\documentclass[twocolumn,amsmath,amssymb,
aps, showpacs,graphicx]{revtex4-2}

\usepackage{graphicx}
\usepackage{dcolumn}
\usepackage{bm}
\usepackage{lineno, hyperref}
\usepackage{natbib}
\begin{document}

\title{Canted antiferromagnetism and excitonic order in gated double-layer graphene}
\author{V. Apinyan} 
\altaffiliation[e-mail:]{v.apinyan@intibs.pl}
\author{T. K. Kope\'{c}} 
\affiliation{Institute of Low Temperature and Structure Research, Polish Academy of Sciences\\ PO. Box 1410, 50-950 Wroc\l{}aw 2, Poland 
}%
%
\begin{abstract}
We study the effects of the electron-electron interactions on the excitonic properties and charge-density modulations in the AB stacked double-layer (DL) graphene, placed in the external gate-potential $V$. The coexistence of the canted antiferromagnetic order and excitonic pairing gap has been studied with the help of the generalized Hubbard model. We calculate the chemical potential $\mu$, the average charge density difference between the layers $\delta{\bar{n}}$, the antiferromagnetic gap-function $\Delta_{\rm AFM}$ and the excitonic order parameters $\Delta_{\sigma}$ in the zero temperature limit. We found that the excitonic pairing order parameter has a larger energy scale than the canted antiferromagnetic gap-function. The charge neutrality, in the DL graphene system, occurs only in the absence of the external gate-potential $V$.
Moreover, we have shown that the values of the antiferromagnetic gap-function $\Delta_{\rm AFM}$ and excitonic order parameter $\Delta_{\sigma}$ are always increasing at the large values of inter-layer Coulomb interaction, while they are decreasing for large values of the applied gate-potential $V$.    
\end{abstract}



\maketitle
%
\section{Introduction}
%
The double-layer (DL) and bilayer graphene (BLG) structures long-time have been the subject of intensive research \cite{cite_1, cite_2, cite_3, cite_4, cite_5, cite_6} due to their extraordinary physical \cite{cite_7, cite_8, cite_9, cite_10} and mechanical properties \cite{cite_11, cite_12}. Particularly, excitonic properties of the BLG structure represent another interesting domain of interest \cite{cite_13, cite_14, cite_15, cite_16, cite_17, cite_18}. The principal achievement in this field was the observation of the large values of the energy gap \cite{cite_19, cite_20, cite_21} and the control of the electronic spectrum of this system \cite{cite_22, cite_23, cite_24, cite_25, cite_26, cite_27, cite_28, cite_29, cite_30, cite_31}. 

When the external electric field is applied to the DL or BLG structures the changes of the carrier charge-density occur in these materials. Those changes lead to the unusual variations of the chemical potential, as a function of electron doping \cite{cite_32}. If the energy cost in the spectrum of single particle excitation is large, then the energy-gap appears in the electronic band structure \cite{cite_33}. Moreover, in some cases, the mentioned gap-function appears also without the external electric field \cite{cite_34}. This situation takes place when considering the electron-electron interaction effects in double-layer structure \cite{cite_13, cite_34} and also when considering the coexistence between the excitonic and antiferromagnetic (AF) orders \cite{cite_35, cite_36, cite_37, cite_38}. The gap-function appearance, with or without electric field, is related first of all to the complicated single-electron reconfiguration effects \cite{cite_39} in the layers of the BLG system. These effects are governed by the inter-layer Coulomb interaction \cite{cite_40} and by the separation distance between the layers \cite{cite_41}, in the DL or BLG systems. More complicated physics is related to the case when considering the competing effects between AF order \cite{cite_31, cite_36} and excitonic effects \cite{cite_31, cite_36, cite_37}. Especially, it has been shown in Refs.\cite{cite_31, cite_36} that the on-site Coulomb interaction stabilizes the AF ordering which, forthermore, gets suppressed at the large values of the bias voltage. Meanwhile, the inter-layer Coulomb repulsion and nonzero voltage stabilize the excitonic order. Here, the principal effect which leads to the formation of the gap-function in the excitation spectrum is related to the formation and  condensation of excitons, which have been the subject of many interesting experimental \cite{cite_39, cite_42, cite_43} and theoretical \cite{cite_35, cite_36, cite_37, cite_38, cite_44, cite_45} investigations, during recent years.    

Beside numerous studies on the problem of coexistence of the antiferromagnetism and excitonic order \cite{cite_35, cite_36, cite_37, cite_38} in the AA BLG and DL structures, there is lack of treatments  concerning double-layer AB or AB BLG. Particularly, the type of the AF order appears to be structurally different in the AB DL systems, due to the canted character of the antiferromagnetism and which is due to the stacking type of the layers in this system. For this reason, the canted antiferromagnetism (CAF) and its coexistence with the excitonic order in the AB DL system is worth of investigations and will be the principal subject of the present paper.       
Here, we study the AB-stacked DL graphene in the presence of the external gate-potential. Moreover, we study the average charge density imbalance between different layers, governed by the external gate-potential and the possible excitonic condensation states. In both layers, we consider different partial-fillings (i.e., fractional average number of particles per-site) of the atomic lattice sites. A particular case of this is accomplished when one has one particle per site in the upper layer, and no particle in the lower layer, i.e, the case of the pumped electron-hole DL graphene. The effects of on-site and inter-layer Coulomb interactions have been treated within the bilayer Hubbard model, and a mean-field analogue theory was constructed to linearize the second order interaction-terms.       

The paper is organized as follows: In Section \ref{sec:Section_2}, we introduce the Hamiltonian of the model. In Section \ref{sec:Section_3}, we obtain the Green function matrix and in Section \ref{sec:Section_4} we give the self-consistent equations. Furthermore, in Section \ref{sec:Section_5} we discuss the results. At the end of the paper, in Section \ref{sec:Section_6}, we give a short conclusion to our paper. The Appendix \ref{sec:Section_7}, is devoted to the calculation of some important coefficients, entering in the series of self-consistent equations.       
%
\section{\label{sec:Section_2} The Hamiltonian of DL graphene}
\begin{figure}
	\includegraphics[scale=0.6]{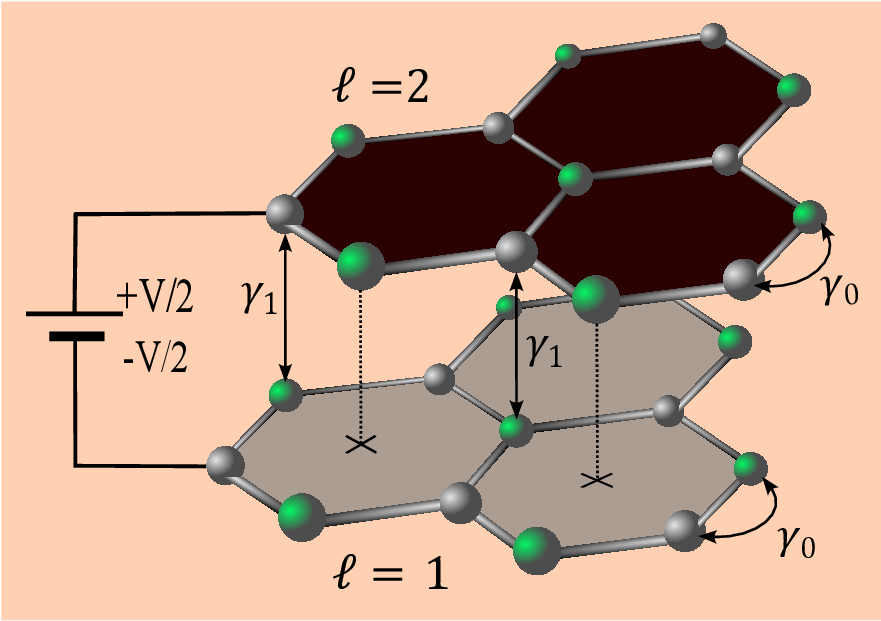}
	\caption{\label{fig:Fig_1}(Color online) The structure of the AB double-layer (DL) graphene in the external electric field potential $V$. The layers of the system have been indicated as $\ell=1$ (the bottom layer) and $\ell=2$ (the upper layer). In the picture, the $A$, $\tilde{A}$ atomic sites are represented by the grey balls, and the $B$,$\tilde{B}$ atomic sites are represented by green balls.}
\end{figure}

%
Our system is described via Hubbard Hamiltonian, written for the two layers    
\begin{eqnarray}
	{\cal{\hat{H}}}_{\rm AB}={\cal{\hat{H}}}_{0}+{\cal{\hat{H}}}_{\rm int}+{\cal{\hat{H}}}_{\rm V},
	\label{Equation_1}
\end{eqnarray}
where ${\cal{\hat{H}}}_{0}$ is the free part, for the non-interacting system, ${\cal{\hat{H}}}_{\rm int}$ is the term which includes the interactions between electrons and ${\cal{\hat{H}}}_{\rm V}$ is Hamiltonian of coupling with the external gate-potential. The free part of the Hamiltonian, in Eq.(\ref{Equation_1}), is    
\begin{eqnarray}	
	{\cal{\hat{H}}}_{0}=&&-\gamma_0\sum_{\left\langle {\bf{r}}{\bf{r}}'\right\rangle}\sum_{\ell\sigma}\left({\hat{a}}^{\dag}_{\ell\sigma}({\bf{r}}){\hat{b}}_{\ell\sigma}({\bf{r}}')+{\rm h.c.}\right)
	\nonumber\\
	&&-\gamma_0\sum_{\left\langle {\bf{r}}{\bf{r}}'\right\rangle}\sum_{\ell\sigma}\left({\hat{\tilde{a}}}^{\dag}_{\ell\sigma}({\bf{r}}){\hat{\tilde{b}}}_{\ell\sigma}({\bf{r}}')+{\rm h.c.}\right)
	\nonumber\\
	&&-\gamma_{1}\sum_{{\bf{r}}\sigma}\left({\hat{b}}^{\dag}_{\sigma}({\bf{r}}){\hat{\tilde{a}}}_{\sigma}({\bf{r}})+{\rm h.c.}\right)
	\nonumber\\
	&&-\mu\sum_{\ell=1,2}\sum_{{\bf{r}}}\hat{n}_{\ell}({\bf{r}}).
	\label{Equation_2}
\end{eqnarray}
The operators ${\hat{a}}_{\ell\sigma}({\bf{r}})$, ${\hat{b}}_{\ell\sigma}({\bf{r}})$, ${\hat{a}}^{\dag}_{\ell\sigma}({\bf{r}})$ and ${\hat{b}}^{\dag}_{\ell\sigma}({\bf{r}})$, in the Eq.(\ref{Equation_2}) are the operators of annihilation and creation for the electrons. The schematic representation of the system in consideration is shown on Fig.~\ref{fig:Fig_1}. The parameter $\gamma_0$ is the hopping amplitude of the electrons in the layers, and the energy parameter $\gamma_1$ describes the hopping of the electrons between the layers. We put for these parameters the values $\gamma_0\sim 3$ eV and $\gamma_1=0.257$ eV (see, Ref.\cite{cite_46}). 	
The symbol $\left\langle ...\right\langle$, in the first two terms, in the Eq.(\ref{Equation _2}), denotes the sum over nearest neighbor lattice site positions. The index $\sigma$, in Eq.(\ref{Equation_2}) describes spins of the electrons, which take two values $\sigma=\uparrow\left(\equiv 1\right) 1,\downarrow\left(\equiv -1\right)$. 
Next, $\mu$ is the chemical potential in the system, coupled with the total electron density operators $\hat{n}_{\ell}$, for the individual layer $\ell$. In turn, the operators $\hat{n}_{\ell}$ are defined in the following way
\begin{eqnarray}
	\hat{n}_{\ell=1}({\bf{r}})=\sum_{\sigma}\hat{a}^{\dag}_{\sigma}({\bf{r}})\hat{a}_{\sigma}({\bf{r}})+\hat{b}^{\dag}_{\sigma}({\bf{r}})b_{\sigma}({\bf{r}}),
	\nonumber\\
	\hat{n}_{\ell=2}({\bf{r}})=\sum_{\sigma}\hat{\tilde{a}}^{\dag}_{\sigma}({\bf{r}})\hat{\tilde{a}}_{\sigma}({\bf{r}})+\hat{\tilde{b}}^{\dag}_{\sigma}({\bf{r}})\hat{\tilde{b}}_{\sigma}({\bf{r}}).
	\label{Equation_3}
\end{eqnarray}
The term ${\cal{\hat{H}}}_{\rm int}$ of Hamiltonian, in Eq.(\ref{Equation_1}), describes the electron-electron interactions in the system   
	\begin{eqnarray}
	{\cal{\hat{H}}}_{\rm int}&=&U\sum_{{\bf{r}}}\sum_{\ell}\hat{n}_{\ell\uparrow}({\bf{r}})\hat{n}_{\ell\eta\downarrow}({\bf{r}})
	\nonumber\\
	&+&W\sum_{{\bf{r}}}\sum_{\sigma\sigma'}\hat{n}_{b\sigma}({\bf{r}})\hat{n}_{\tilde{a}\sigma'}({\bf{r}}),
	\label{Equation_4}
\end{eqnarray}
where $U$ is the on-site Hubbard interaction and $W$ is the Coulomb interaction potential between the layers, which has also local charachter, as the potential $U$. 
The coupling with the external electric field $V$ is described by Hamiltonian ${\cal{\hat{H}}}_{\rm V}$ 
\begin{eqnarray}
	&&{\cal{\hat{H}}}_{\rm V}=\frac{V}{2}\sum_{{\bf{r}}}\left(\hat{n}_{2}\left({\bf{r}}\right)-\hat{n}_{1}\left({\bf{r}}\right)\right).
	\label{Equation_5}
\end{eqnarray}
%
%
%
%
We suppose that the potential of the electric field at the upper layer (with the layer index $\ell=2$) is $+\frac{V}{2}$ and the potential of the electric field at the bottom layer (with the layer index $\ell=1$) is $-\frac{V}{2}$ . This is represented in Fig.~\ref{fig:Fig_1}. 
%
\section{\label{sec:Section_3} Hartree-Fock decoupling and order parameters}
%
To proceed, we pass to the Grassmann representation for the fermions (where the operators ${\hat{a}}_{\ell\sigma}({\bf{r}})$, ${\hat{b}}_{\ell\sigma}({\bf{r}})$, ${\hat{a}}^{\dag}_{\ell\sigma}({\bf{r}})$ and ${\hat{b}}^{\dag}_{\ell\sigma}({\bf{r}})$ are replaced by the complex numbers ${a}_{\ell\sigma}({\bf{r}})$, ${b}_{\ell\sigma}({\bf{r}})$, ${\bar{a}}_{\ell\sigma}({\bf{r}})$ and ${\bar{b}}_{\ell\sigma}({\bf{r}})$) and we linearize the biquadratic density terms using the Hubbard-Stratanovich transformation. 
The total action of the AB DL graphene system could be written as
\begin{eqnarray}
	{\cal{S}}=\sum_{\eta}{\cal{S}}_{B}\left[\bar{\eta},\eta\right]+\int^{\beta}_{0}d\tau {\hat{\cal{H}}}_{\rm AB}(\tau), 
	\label{Equation_6}
\end{eqnarray}
where $S_{B}\left[\bar{\eta},\eta\right]$, in the first term, in the Eq.(\ref{Equation_6}), is the Berry action, which is defined as
\begin{eqnarray}
	{\cal{S}}_{B}\left[\bar{\eta},\eta\right]=\sum_{{\bf{r}}\sigma}\int^{\beta}_{0}d\tau \bar{\eta}_{\sigma}({\bf{r}}\tau)\partial_{\tau}\eta_{\sigma}({\bf{r}}\tau)
	\label{Equation_7}
\end{eqnarray}
and the summation index $\eta$ indicates the type of the particles, i.e.,
\begin{eqnarray}
	\footnotesize
	\arraycolsep=0pt
	\medmuskip = 0mu
	\eta
	=\left\{
	\begin{array}{cc}
		\displaystyle   a,b \ \ \  $if$ \ \ \ \ell=1,
		\newline\\
		\newline\\
		{\footnotesize
			\arraycolsep=0pt
			\medmuskip = 0mu
			\begin{array}{cc}
				\displaystyle  & \tilde{a},\tilde{b} \ \ \  $if$ \ \ \  \ell=2.
		\end{array}}
	\end{array}\right.
	\label{Equation_8}
\end{eqnarray}
Next $\beta=1/k_{B}T$ where $T$ is the temperature, and the integration variable $\tau$ is the imaginary-time $\tau$, given from the interval $0\leq\tau\leq\beta$. 

Furthermore, we use the fermionic path integral approach, already employed in Ref.\cite{cite_31}. The linearization procedure for the bilinear fermionic density terms is quite involved \cite{cite_31}, and, for this reason, we give, here, only the form of the resulting action. Hubbard-Stratanovich decoupling, for density-density terms, in the Hamiltonian in Eq.(\ref{Equation_4}), results in Hartree-Fock-like expressions for the excitonic gap parameter $\Delta_{\sigma}$ and antiferromagnetic gap-function $\Delta_{\rm AFM}$. In particular, the antiferromagnetic gap-function $\Delta_{\rm AFM}$ for the $\eta$-type sublattice site is 
\begin{eqnarray}
\Delta^{\eta}_{\rm AFM}=\frac{U}{2}\left\langle n_{\eta\uparrow}-n_{\eta\downarrow}\right\rangle. 
\label{Equation_9}
\end{eqnarray}
Assuming the staggered CAF order in the system, we can write 
\begin{eqnarray}
\nonumber\\
\Delta^{a}_{\rm AFM}&=&-\Delta_{\rm AFM},
\nonumber\\ 
\Delta^{\tilde{a}}_{\rm AFM}&=&-\Delta_{\rm AFM},
\nonumber\\
\Delta^{b}_{\rm AFM}&=&\Delta_{\rm AFM}, 
\nonumber\\
\Delta^{\tilde{b}}_{\rm AFM}&=&\Delta_{\rm AFM}. 
\label{Equation_10}
\end{eqnarray}
We assume, in general, that $|\Delta^{\tilde{a}}_{\rm AFM}|=|\Delta^{b}_{\rm AFM}|$, because the corresponding electron densities are localized, in the layers, on the same lattice site positions in the $x-y$ plane. The case when $\Delta^{\tilde{a}}_{\rm AFM}\neq\Delta^{b}_{\rm AFM}$ (representing the case of the inhomogeneous CAF order) is out of the scope of discussion in this paper.  
Thus, we have the following averages
\begin{eqnarray}
\Delta_{\rm AFM}&=&\frac{U}{2}\left\langle n_{b\uparrow}-n_{b\downarrow}\right\rangle,
\nonumber\\
\Delta_{\sigma}&=&W\left\langle \bar{b}_{\sigma}({\bf{r}}\tau)\tilde{a}_{\sigma}({\bf{r}}\tau)\right\rangle.
\label{Equation_11}
\end{eqnarray}
The averages $\left\langle ... \right\rangle$, in Eq.(\ref{Equation_11}), can be expressed with the help of the partition function of the system
\begin{eqnarray}
\left\langle ... \right\rangle=\frac{1}{Z}\int\left[{\cal{D}}\bar{\psi}{\cal{D}}\psi\right]...e^{-{\cal{S}}\left[\bar{\psi},\psi\right]} ,
\label{Equation_12}
\end{eqnarray}
where $Z$ is the partition function of the AB DL system 
\begin{eqnarray}
	Z=\int\left[{\cal{D}}\bar{\psi}{\cal{D}}\psi\right]e^{-{\cal{S}}\left[\bar{\psi},\psi\right]}.
	\label{Equation_13}
\end{eqnarray}
We have introduced in the Eqs.(\ref{Equation_12}) and (\ref{Equation_13}) Nambu spinors $\bar{\psi}_{\sigma}({\bf{k}}\nu_{n})$ and their conjugates fields $\left(\bar{\psi}_{\sigma}({\bf{k}}\nu_{n})\right)^{T}$, as
\begin{eqnarray} 
	\left(\bar{\psi}_{\sigma}({\bf{k}}\nu_{n})\right)^{T}=\left(\bar{a}_{\sigma}({\bf{k}}\nu_{n}),\bar{b}_{\sigma}({\bf{k}}\nu_{n}),\bar{\tilde{a}}_{\sigma}({\bf{k}}\nu_{n}),\bar{\tilde{b}}_{\sigma}({\bf{k}}\nu_{n})\right),
	\nonumber\\
	\label{Equation_14}
\end{eqnarray}
where $\nu_{n}$ are Matsubara frequencies for fermionic Grassmann field $\nu_{n}=\pi(2n+1)/\beta$, with $n=0,\pm1,\pm2,...$. The action in the Eq.(\ref{Equation_13}) is represented in the following form
\begin{eqnarray} 
	&&{\cal{S}}\left[\bar{\psi},\psi\right]=\frac{1}{\beta{N}}\sum_{{\bf{k}}\nu_{n}\sigma}\bar{\psi}_{\sigma}({\bf{k}}\nu_{n})\hat{{\cal{G}}}^{-1}_{\sigma}({\bf{k}}\nu_{n}){\psi}_{\sigma}({\bf{k}}\nu_{n}),
	\nonumber\\
	\label{Equation_15}
\end{eqnarray}
where $\hat{{\cal{G}}}^{-1}_{\sigma}({\bf{k}}\nu_{n})$ is the inverse Green's function matrix
\newline
\begin{equation}
	\noindent 
	\resizebox{\linewidth}{!}{\columnsep=1pt %
		$\hat{{\cal{G}}}^{-1}_{\sigma}({\bf{k}}\nu_{n})=
		\left(\begin{matrix}
			-\mu_{1\sigma}-i\nu_{n} & -\tilde{\gamma}_{1{\bf{k}}} & 0 & 0\\
			-\tilde{\gamma}_{2{\bf{k}}} & -\mu_{2\sigma}-i\nu_{n} & -{\gamma}_1-\bar{\Delta}_{\sigma} &  0\\
			0 & -{\gamma}_1-{\Delta}_{\sigma} & -\mu_{3\sigma}-i\nu_{n} & -\tilde{\gamma}_{2{\bf{k}}} \\
			0 & 0 & -\tilde{\gamma}_{1{\bf{k}}} & -\mu_{4\sigma}-i\nu_{n} 
		\end{matrix}\right)$}.
	\label{Equation_16}
\end{equation}
Here, we have introduced the effective chemical potentials $\mu_{i}$ with $i=1,...,4$, for different spin directions
\begin{eqnarray}
	\mu_{i\sigma}&=&\mu+(-1)^{i}\sigma\Delta_{\rm AFM}+\frac{V}{2}-\frac{U}{2}\bar{n}_{b}
	\nonumber\\
	&&+\left(-1\right)^{i+1}\left(i-1\right)2\sigma{W},
	\nonumber\\
	\mu_{j\sigma}&=&\mu+(-1)^{i}\sigma\Delta_{\rm AFM}-\frac{V}{2}-\frac{U}{2}\bar{n}_{\tilde{a}},
	\label{Equation_17}
\end{eqnarray}
where $i=1,2$ and $j=3,4$. 
The chemical potentials, in the Eq.(\ref{Equation_17}), express the effective, single-particle, excitation spectrum in our system, composed of four sublattices. They contain the Coulomb interactions $U$, $W$, the external gate-potential $V$ and the CAF order parameter $\Delta_{\rm AFM}$.  
The parameters $\tilde{\gamma}_{i{\bf{k}}}$ with $i=1,2$, in Eq.(\ref{Equation_16}), are the dispersion relations in the layers of the AB DL graphene 
\begin{eqnarray}
\tilde{\gamma}_{1{\bf{k}}}=\tilde{\gamma}_{0}\left(e^{-ik_{x}}+2e^{i\frac{k_{x}}{2}}\cos(k_{y}\sqrt{3})\right),
\nonumber\\
\tilde{\gamma}_{2{\bf{k}}}=\tilde{\gamma}_{0}\left(e^{ik_{x}}+2e^{-i\frac{k_{x}}{2}}\cos(k_{y}\sqrt{3})\right).
\label{Equation_18}
\end{eqnarray}
We realize that $\tilde{\gamma}_{2{\bf{k}}}=\tilde{\gamma}^{\ast}_{1{\bf{k}}}$. 
In the next section, we derive the explicit form of the set of self-consistent equations. 
%
\section{\label{sec:Section_4} The Self-consistent equations}
%
Next, we use the partition function in the Eq.(\ref{Equation_13}), to calculate the average charge densities $\bar{n}_{b}$ and $\bar{n}_{\tilde{a}}$ at the sublattice sites $B$ and $\tilde{A}$
\begin{eqnarray}
	\bar{n}_{b}=\sum_{\sigma}\left\langle \bar{b}_{\sigma}({\bf{r}}\tau)b_{\sigma}({\bf{r}}\tau)\right\rangle, 
	\nonumber\\
	\bar{n}_{\tilde{a}}=\sum_{\sigma}\left\langle \bar{\tilde{a}}_{\sigma}({\bf{r}}\tau)\tilde{a}_{\sigma}({\bf{r}}\tau)\right\rangle.
	\label{Equation_19}
\end{eqnarray}
Now, we will define the equations, which describe the average particle-filling numbers (with the coefficient $1/\kappa$) and the average particle-density difference (which we note by $\delta{\bar{n}}$) between the layers. Those average number of particles concern the sublattices $B$ (in the bottom layer $\ell=1$) and $\tilde{A}$ (in the upper layer $\ell=2$). Those equations are
\begin{eqnarray}
&&\bar{n}_{b}+\bar{n}_{\tilde{a}}=\frac{1}{\kappa},
\nonumber\\
&&\bar{n}_{\tilde{a}}-\bar{n}_{b}=\frac{\delta{\bar{n}}}{2}.
\label{Equation_20}
\end{eqnarray}
Then, we can calculate the averages $\bar{n}_{b}$ and $\bar{n}_{\tilde{a}}$ with the help of $\kappa$ and $\delta{\bar{n}}$
\begin{eqnarray}
	\bar{n}_{b}=\frac{1}{2}\left(\frac{1}{\kappa}-\frac{\delta{\bar{n}}}{2}\right),
	\nonumber\\
	\bar{n}_{\tilde{a}}=\frac{1}{2}\left(\frac{1}{\kappa}+\frac{\delta{\bar{n}}}{2}\right).
	\label{Equation_21}
\end{eqnarray}
%
%
\begin{figure}
	\includegraphics[scale=0.26]{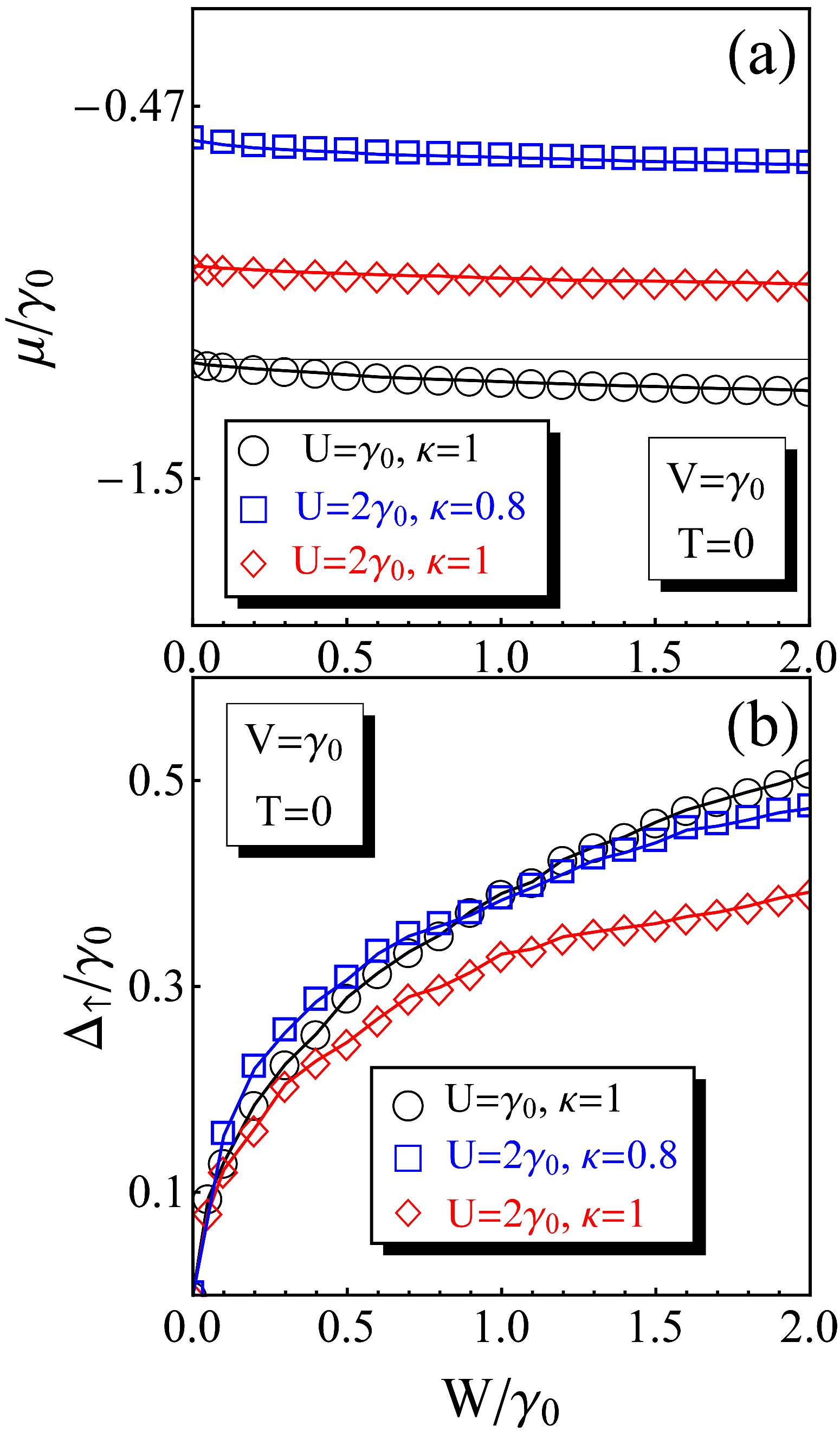}
	\caption{\label{fig:Fig_2}(Color online) The numerical results for $\mu$ (see panel (a)), $\Delta_{\uparrow}$ (see in panel (b)), as a function of the inter-layer Coulomb interaction parameter $W$. The calculations have been done in the zero temperature limit, and different values of the interaction parameter $U$ and inverse-filling coefficient $\kappa$ have been considered. The external gate-potential has been fixed at the value $V=\gamma_0=3$ eV.}
\end{figure} 
%
%
%
\begin{figure}
	\includegraphics[scale=0.26]{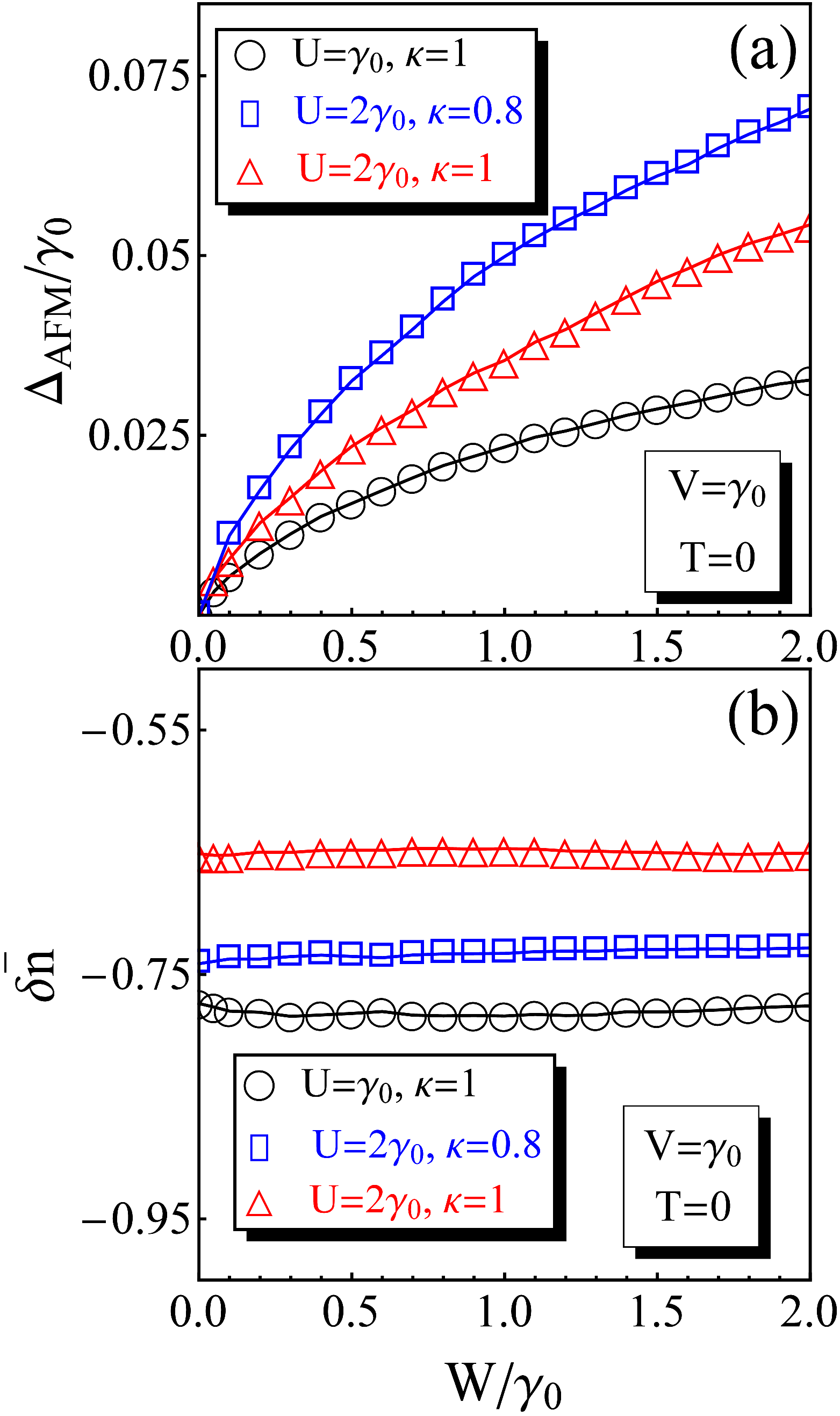}
	\caption{\label{fig:Fig_3}(Color online) The numerical results for $\Delta_{\rm AFM}$ (see in panel (a)) and $\delta{\bar{n}}$ (see in panel (b)) as a function of the inter-layer Coulomb interaction parameter $W$. The calculations have been done in the zero temperature limit, and different values of the interaction parameter $U$ and inverse-filling coefficient $\kappa$ have been considered. The external gate-potential has been fixed at the value $V=\gamma_0=3$ eV.}
\end{figure} 
%
%
We evaluate the statistical averages in Eqs.(\ref{Equation_11}) and (\ref{Equation_19}) by using the inverse Green function matrix, given in Eq.(\ref{Equation_16}). Furthermore, we combine the definitions in the Eq.(\ref{Equation_11}) with those in Eq.(\ref{Equation_20}), to obtain the complete set of self-consistent equations. We use also the fact that $\bar{n}_{a}=\bar{n}_{b}$ and $\bar{n}_{\tilde{a}}=\bar{n}_{\tilde{b}}$. The system of self-consistent equations for AB DL system reads:
\begin{eqnarray}
	\frac{1}{\kappa}&=&-\frac{1}{N}\sum_{i{\bf{k}},\sigma}\alpha_{i{\bf{k}}\sigma}n_{F}(\mu-\varepsilon_{i\sigma}({\bf{k}})),
	\nonumber\\
	\frac{\delta{\bar{n}}}{2}&=&-\frac{1}{N}\sum_{i{\bf{k}},\sigma}\beta_{i{\bf{k}}\sigma}n_{F}(\mu-\varepsilon_{i\sigma}({\bf{k}})),
	\nonumber\\
	\Delta_{\rm AFM}&=&-\frac{U}{2N}\sum_{i{\bf{k}},\sigma}\sigma\gamma_{i{\bf{k}}\sigma}n_{F}(\mu-\varepsilon_{i\sigma}({\bf{k}})),
	\nonumber\\
	\Delta_{\sigma}&=&\frac{W}{N}\sum_{i{\bf{k}}}\delta_{i{\bf{k}}\sigma}n_{F}(\mu-\varepsilon_{i\sigma}({\bf{k}})).
	\label{Equation_22}
\end{eqnarray}
The explicit forms of the coefficients $\alpha_{i{\bf{k}}\sigma}$, $\beta_{i{\bf{k}}\sigma}$, $\gamma_{i{\bf{k}}\sigma}$ and $\delta_{i{\bf{k}}\sigma}$, in Eq.(\ref{Equation_22}), are given in Appendix \ref{sec:Section_7}. Next, the function $n_{F}(x)$, in Eq.(\ref{Equation_22}), is the Fermi-Dirac distribution
\begin{eqnarray}
	n_{F}\left(x\right)=\frac{1}{e^{\beta{(x-\mu)}}+1}.
	\label{Equation_23}
\end{eqnarray}
Furthermore, the energy parameters $\varepsilon_{i\sigma}({\bf{k}})$, in Eq.(\ref{Equation_22}), define the energy band-structure in the AB DL graphene. They are given as
\begin{eqnarray}
	\varepsilon_{i\sigma}({\bf{k}})=\bar{\mu}_{1\sigma}+\frac{\left(-1\right)^{i}}{2}\sqrt{A_{\bf{k}\sigma}-2\sqrt{B_{{\bf{k}}\sigma}}},
	\nonumber\\
	\varepsilon_{j\sigma}({\bf{k}})=\bar{\mu}_{2\sigma}+\frac{\left(-1\right)^{j}}{2}\sqrt{A_{\bf{k}\sigma}+2\sqrt{B_{{\bf{k}}\sigma}}},
	\label{Equation_24}
\end{eqnarray}
where $i=1,2$ and $j=3,4$, and
\begin{eqnarray}
\bar{\mu}_{1\sigma}=\left({\mu}_{1\sigma}+{\mu}_{2\sigma}\right)/2,
\nonumber\\
\bar{\mu}_{2\sigma}=\left({\mu}_{3\sigma}+{\mu}_{4\sigma}\right)/2.
	\label{Equation_25}
\end{eqnarray}
The energy and spin-dependent parameters $A_{{\bf{k}}\sigma}$ and $B_{{\bf{k}}\sigma}$, in the Eq.(\ref{Equation_24}), have been obtained as
\begin{eqnarray}
A_{\bf{k}\sigma}&=&2|\Delta_{\sigma}+\gamma_1|^{2}+4|\tilde{\gamma}_{\bf{k}}|^{2}+4|\Delta_{\rm AFM}|^{2}+\left(\mu_{1\sigma}-\mu_{2\sigma}\right)^{2},
\nonumber\\
B_{\bf{k}\sigma}&=&|\Delta_{\sigma}+\gamma_1|^{4}+4|\Delta_{\sigma}+\gamma_1|^{2}|\tilde{\gamma}_{\bf{k}}|^{2}
\nonumber\\
&&+4\left(\mu_{1\sigma}-\mu_{2\sigma}\right)|\Delta_{\sigma}+\gamma_1|^{2}\Delta_{\rm AFM}
\nonumber\\
&&+4\left(\mu_{1\sigma}-\mu_{2\sigma}\right)^{2}\left(|\tilde{\gamma}_{\bf{k}}|^{2}+|\Delta_{\rm AFM}|^{2}\right).
\label{Equation_26}
\end{eqnarray}
In the following, we solve numerically the system of equations in the Eq.(\ref{Equation_22}) with the help of Newton's fast convergent algorithm. We will discuss the obtained results in the next section.
%
\section{\label{sec:Section_5} Results}
%
Here, we will discuss the main results obtained in the present paper. In Fig.~\ref{fig:Fig_2}, we calculate numerically the chemical potential $\mu$ and the excitonic order parameter $\Delta_{\uparrow}$. In panel (a), the chemical potential is presented as a function of the inter-layer Coulomb interaction potential $W$. Several values of the on-site Coulomb potential $U$ (see in Eq.(\ref{Equation_4})) have been considered in the picture. Different average particle filling regimes have been used. All results have been done in the zero temperature limit $T=0$, and the external gate-potential was set at the value $V=\gamma_0$. The black, red curves show the $\mu\left(W\right)$ dependence for $U=\gamma_0$, $U=2\gamma_0$, and correspond to the case $\kappa=1$. The blue-squared points correspond to the case of the smaller value of the coefficient $\kappa$, namely $\kappa=0.8$. The value $\kappa=1$ corresponds to the case when the AB DL was initially pumped with the electrons in the upper layer and the holes in the lower layer, or vice versa. We see that $\mu$ is nearly constant as a function of $W$. The negative values of $\mu$ indicate on the possibility of the existence of the stable excitonic condensate state \cite{cite_34}. The same, nearly constant, behavior was observed for the average charge density imbalance function $\delta\bar{n}$ (see in panel (b), in Fig.~\ref{fig:Fig_3}). We observe also that the values of the obtained chemical potential are smaller (with modulus) for the smaller value of the coefficient $\kappa$, namely for $\kappa=0.8$ (see the blue curve on panel (a), in Fig.~\ref{fig:Fig_2}). This means that the energy cost for the creation of the single-particle excitation is lower in that case. We see in panel (b), in Fig.~\ref{fig:Fig_3}, that the values of the function $\delta\bar{n}$ are negative, for all considered values of the potential $U$. This fact shows that the system is in the regime of the hole-electron DL, i.e., there are fewer electrons in the upper layer than in the bottom layer ($\bar{n}_{2}<\bar{n}_{1}$). The calculated excitonic order parameter $\Delta_{\uparrow}$, and the CAF gap-function $\Delta_{\rm AFM}$ have been showed respectively on panel (b) in the Fig.~\ref{fig:Fig_2} and on panel (a), in Fig.~\ref{fig:Fig_3}. We see that they both are monotonically increasing functions of $W$, i.e., they are continuously increasing with $W$, for all values of the localizing Coulomb potential $U$.
%
%
\begin{figure}
	\includegraphics[scale=0.26]{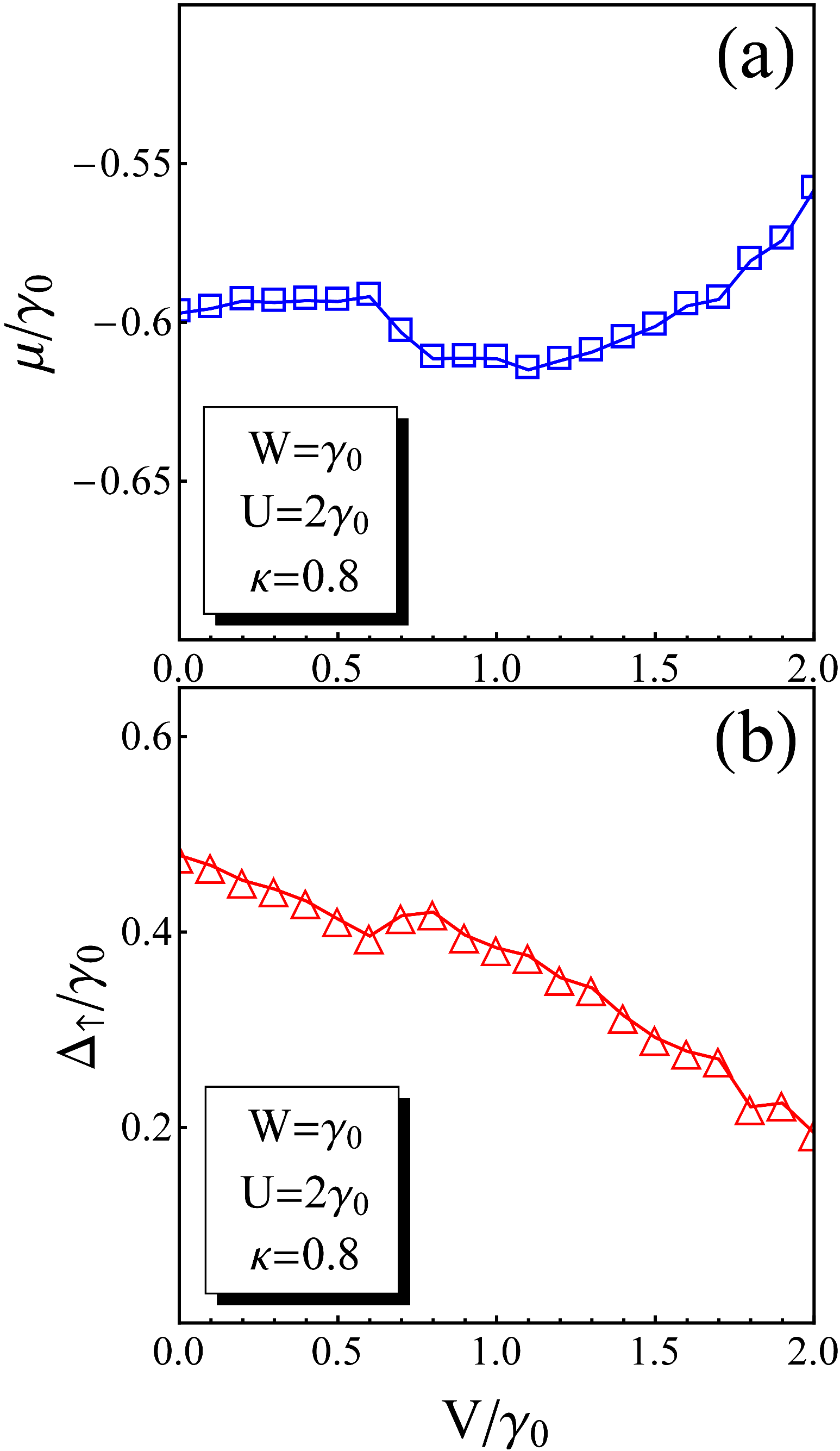}
	\caption{\label{fig:Fig_4}(Color online) The numerical results for $\mu$ (see on panel (a)) and $\Delta_{\uparrow}$ (see on panel (b)), as a function of the gate-potential $V$. The calculations have been done in the zero temperature limit $T=0$. The Coulomb interaction parameter $U$, the inter-layer interaction $W$ and the inverse filling-coefficient $\kappa$ have been set respectively at the values $U=2\gamma_0$, $W=\gamma_0$ and $\kappa=0.8$.}
\end{figure}
%
%
In addition, we see that the excitonic and CAF order parameters get larger values in the case of the small value of the inverse filling coefficient $\kappa$, notably, for $\kappa=0.8$. This is directly related to the fact that the chemical potential $\mu$ takes small values in that case (see blue-square points on panel (a), in Fig.~\ref{fig:Fig_2}), i.e., when the energy cost for single-particle excitations in the system is lower.

\begin{figure}
	\includegraphics[scale=0.26]{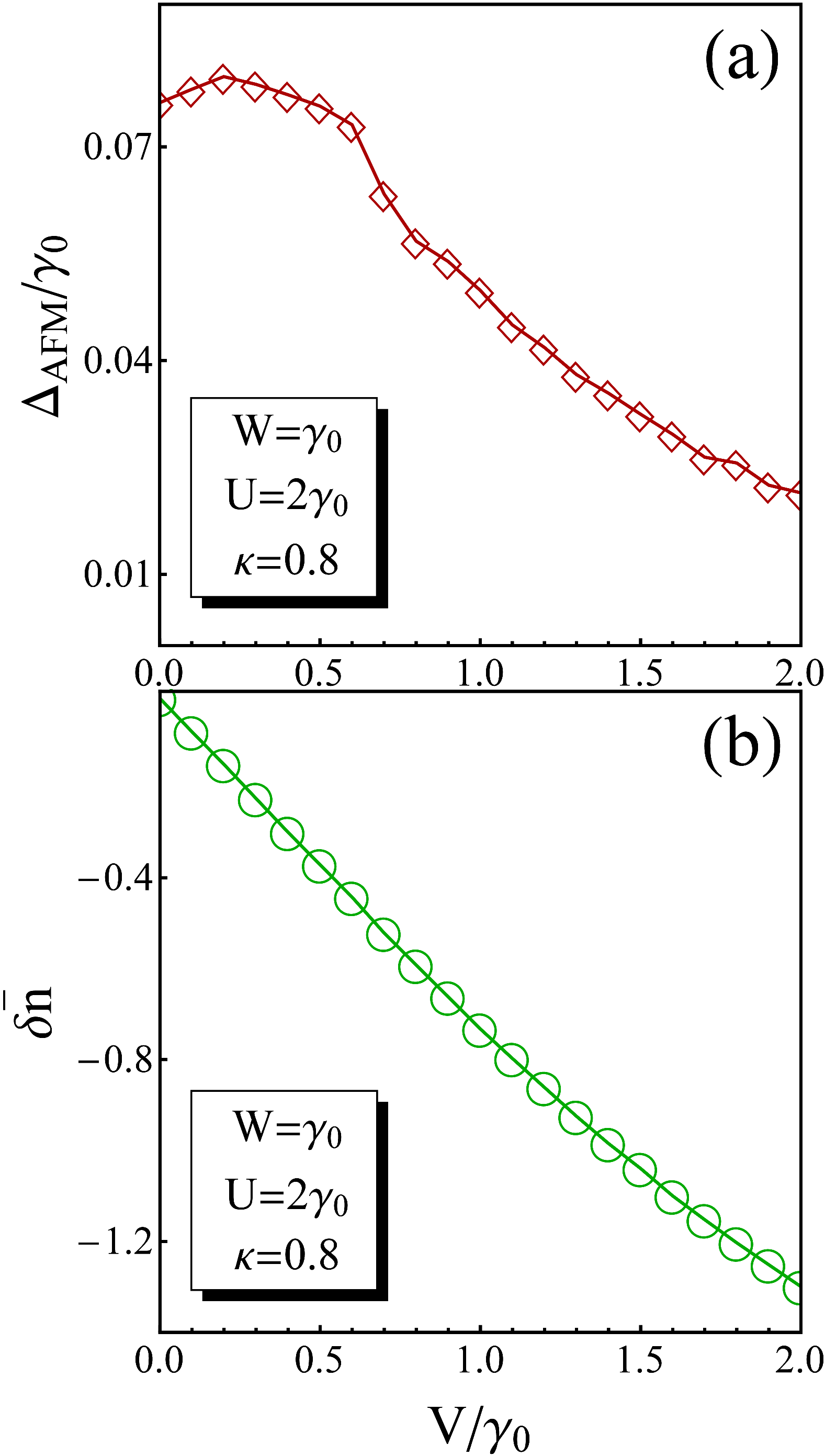}
	\caption{\label{fig:Fig_5}(Color online) The numerical results for $\Delta_{\rm AFM}$ (see on panel (a)) and $\delta{\bar{n}}$ (see on panel (b)), as a function of the gate-potential $V$. The calculations have been done in the zero temperature limit $T=0$. The Coulomb interaction parameter $U$, the inter-layer interaction $W$ and the inverse filling-coefficient $\kappa$ have been set respectively at the values $U=2\gamma_0$, $W=\gamma_0$ and $\kappa=0.8$.}
\end{figure}

Furthermore, in Figs.~\ref{fig:Fig_4} and ~\ref{fig:Fig_5}, we have shown the $V$-dependence of the same physical quantities. The results, in Figs.~\ref{fig:Fig_4} and ~\ref{fig:Fig_5}, have been done for the case $W=\gamma_0$, $U=2\gamma_0$ and $\kappa=0.8$. The calculations were performed in the zero temperature limit. We see on panel (a) of Fig.~\ref{fig:Fig_4} that $\mu<0$, for all considered values of the applied gate potential $V$. The excitonic order parameter and the CAF deviation function are no longer open functions (see on panels (b) and (a), in Figs.~\ref{fig:Fig_4} and ~\ref{ fig:Fig_5}, respectively). Particularly, they are decreasing when the one increase the gate-potential parameter $V$, from zero up to the value $V=2\gamma_0$. The general strategy for finding the dependence on $V$ for the excitonic gap function is to fix the distance between the layers, which, therefore, will lead to the fixed numerical value for the interlayer Coulomb interaction $W$ and for Hubbard- $U$ local interaction. Those values have been show on all panels in Figs.~\ref{fig:Fig_4} and ~\ref{fig:Fig_5}. Namely, we have $W=\gamma_0=3$ eV and $U=2\gamma_0=6$ eV. The average charge density imbalance $\delta{\bar{n}}$ shown on panel (b) in Fig.~\ref{fig:Fig_5} decreases dramatically with $V$, being always negative, i.e. $\delta {\bar{n}}\left(V\right)<0$ ($\bar{n}_{2}<\bar{n}_{1}$). Moreover, for large values of $V$ (see the values of $V$ in $\gamma_0\leq V\leq 2\gamma_0$), the average charge imbalance between the layers becomes very large and the layer with $\ell=2$ becomes depopulated from electrons. Thus, by increasing the external potential we reconfigure the electrons distributions in the layers and the upper layer with $\ell=2$ becomes pumped from the electrons. We observe also that the exact charge neutrality ($\delta{\bar{n}}=0$) occurs only at the value $V=0$ of the external gate-potential.               
%
%
%
%
\section{\label{sec:Section_6} Discussions and Conclusion}
%
We have studied the coexistence of the excitonic and antiferromagnetic orders in the double-layer AB-stacked graphene. The external gate-voltage was applied to the DL structure. The principal attention has been put on the Coulomb interaction effects in the bilayer system, which have been treated with the help of the bilayer Hubbard Hamiltonian. In addition, the electronic energy spectrum was obtained using effective chemical potentials. The partial filling has been considered in this article, i.e. when the average number of electrons at the given lattice site position is less than one.

The numerical calculations have been performed by using Newton's fast convergent algorithm, and the results have been obtained at the zero temperature limit.  
The calculated negative values of the chemical potential in the system show the possibility of the existence of the excitonic condensate state in the DL system, which can be tuned by varying the separation distance between the layers in the DL structure.
We have shown that the energy scales of the excitonic order parameter $\Delta_{\uparrow}$ are much larger than the energy scales corresponding to the CAF gap, i.e. $\Delta_ {\uparrow}\gg\Delta_{\rm AFM }$.This finding is just the opposite of the result in Ref.\cite{cite_31}, where it has been shown that the AF gap is much larger than the excitonic order parameter, i.e., $\Delta_{\rm AFM}\gg\Delta_{\uparrow}$. 
Moreover, the excitonic order parameter survives only for one spin direction, particularly for $\sigma=\uparrow$.

Another interesting observation is that the average charge imbalance $\delta{\bar{n}}$ is always negative for all values of the Coulomb interaction between the layers and of the electric field potential which are given in the intervals $0 \leq W\leq 2\gamma_{0 }$ and $0\leq V\leq 2\gamma_0$. Therefore, the average number of electrons in the upper layer is always less than in the lower layer. The AB DL system, with this type of configuration, could be interesting for modern technological applications \cite{cite_46, cite_47, cite_48, cite_49, cite_50} and is more useful thanks to the high speed of electrons in the graphene layers \cite {cite_51, cite_52 , cite_53}. The changes of the charge imbalance with the applied gate-potential show the possibility to tune the AB DL system from the electron-electron to electron-hole type, when increasing the external gate-potential $V$. The exact charge-neutrality, i.e., when $\delta{\bar{n}}=0$, occurs only at the zero value of the external voltage $V=0$. The charge imbalance is stronger for large values of $V$, which means that large values of $V$ effectively stabilize the electron-hole type bilayer rather than the electron-electron one. For fixed $V$, the variation of $\delta{\bar{n}}$, as a function of $W$, is almost constant, so it appears that the average charge imbalance is indifferent to changes in the distance of separation between layers.
Moreover, all non-zero values of the gate potential have a destructive effect on the excitonic gap parameter, while the antiferromagnetic gap function increases, at relatively small values of $V$, especially when $0\leq V \leq 0.2\gamma_0$. 

To summarize, the chemical potential and the charge density imbalance between the layers are constant linear functions of the Coulomb potential between the layers and therefore they do not depend on the charge modulations in the layers and the separation distance between layers. The excitonic and CAF order parameters are increasing functions of the interaction parameter. The applied gate potential changes the physical parameters more significantly than the Coulomb interaction between the layers (note that the Coulomb potential between the layers varies when changing the separation distance between the layers). We have shown that the energy scales, corresponding to the CAF order parameter, are much smaller than the excitonic energy scales and we attribute this result to the stacking type of the system, i.e., the AB stacking of the layers. It is worth noting that in the case of AA-type stacked DL graphene, the result is exactly the opposite \cite{cite_31} and the opposite effect occurs for the mentioned energy scales.

We think that the results, obtained in the paper, could be important from both theoretical and experimental points of view. In particular, the energy scales of the antiferromagnetic and excitonic orders were unexpectedly modified when considering the effect of CAF in the AB stacked double layer, instead of AA type of stacking (see, in comparison, the results in \cite{cite_31}).
The results in the paper could be also helpful and postponed for the applications of the AB DL system in modern electronics, optoelectronics and photonics technologies as a device, where the electron density reconfigurations and optical properties get simultaneous changes. 
\appendix 
%
\section{\label{sec:Section_7} The calculation of important coefficients}
%
In this Section, we show the explicit forms of the coefficients-parameters entering in the right-hand sides of equations in Eq.(\ref{Equation_22}). 
Particularly, we have

\begin{eqnarray}
\footnotesize
\arraycolsep=0pt
\medmuskip = 0mu
\alpha_{i{\bf{k}}\sigma}
=\left\{
\begin{array}{cc}
\displaystyle  & \frac{\left(-1\right)^{i+1}}{\varepsilon_{1\sigma}({\bf{k}})-\varepsilon_{2\sigma}({\bf{k}})}\prod_{j=3,4}\frac{{\cal{P}}^{\left(3\right)}\left(\varepsilon_{i\sigma}({\bf{k}})\right)}{\left(\varepsilon_{i\sigma}({\bf{k}})-\varepsilon_{j\sigma}({\bf{k}})\right)}, \ \ \ $if$ \ \ \  i=1,2.
\newline\\
\newline\\
\displaystyle  & \frac{\left(-1\right)^{i+1}}{\varepsilon_{3\sigma}({\bf{k}})-\varepsilon_{4\sigma}({\bf{k}})}\prod_{j=1,2}\frac{{\cal{P}}^{\left(3\right)}\left(\varepsilon_{i\sigma}({\bf{k}})\right)}{\left(\varepsilon_{i\sigma}({\bf{k}})-\varepsilon_{j\sigma}({\bf{k}})\right)}, \ \ \ $if$ \ \ \  i=3,4,
\end{array}\right.
\label{Equation_A_1}
\end{eqnarray}
where the band-structure quasienergies $\varepsilon_{i\sigma}({\bf{k}})$ (with $i=1,...,4$) have been obtained at the end of the Section \ref{sec:Section_4} and third order polynomial ${\cal{P}}^{\left(3\right)}(x)$ in Eq.(\ref{Equation_A_1}) is defined as 
\begin{eqnarray}
	&&{\cal{P}}^{\left(3\right)}\left(x\right)=2x^{3}+a_{1\sigma}x^{2}+b_{1\sigma}x+c_{1\sigma}.
	\label{Equation_A_2}
\end{eqnarray}
Here, 
\begin{eqnarray}
	a_{1\sigma}&=&-3\left(\bar{\mu}_{1\sigma}+\bar{\mu}_{2\sigma}\right)
	\nonumber\\
	b_{1\sigma}&=&-2|\tilde{\gamma}_{\bf{k}}|^{2}-2\Delta^{2}_{\rm AFM}+2\sigma\left(\bar{\mu}_{1\sigma}-\bar{\mu}_{2\sigma}\right)\Delta_{\rm AFM}
	\nonumber\\
	&&+\left(\bar{\mu}_{1\sigma}+\bar{\mu}_{2\sigma}\right)^{2}+2\bar{\mu}_{1\sigma}\bar{\mu}_{2\sigma},
	\nonumber\\
	c_{1\sigma}&=&\left(\bar{\mu}_{1\sigma}+\bar{\mu}_{2\sigma}\right)|\tilde{\gamma}_{\bf{k}}|^{2}+\left(\bar{\mu}_{1\sigma}+\bar{\mu}_{2\sigma}\right)\Delta^{2}_{\rm AFM}
	\nonumber\\
	&&+\sigma\left(\bar{\mu}^{2}_{2\sigma}-\bar{\mu}^{2}_{1\sigma}\right)-\bar{\mu}_{1\sigma}\bar{\mu}_{2\sigma}\left(\bar{\mu}_{1\sigma}+\bar{\mu}_{2\sigma}\right).
	\label{Equation_A_3}
\end{eqnarray}
Next, in the second of equations in the system in the Eq.(\ref{Equation_22}), we get for the coefficient $\beta_{i{\bf{k}}\sigma}$:
\begin{eqnarray}
\footnotesize
\arraycolsep=0pt
\medmuskip = 0mu
\beta_{i{\bf{k}}\sigma}
=\left\{
\begin{array}{cc}
\displaystyle  & \frac{\left(-1\right)^{i+1}}{\varepsilon_{1\sigma}({\bf{k}})-\varepsilon_{2\sigma}({\bf{k}})}\prod_{j=3,4}\frac{{\cal{P}}^{\left(2\right)}\left(\varepsilon_{i\sigma}({\bf{k}})\right)}{\left(\varepsilon_{i\sigma}({\bf{k}})-\varepsilon_{j\sigma}({\bf{k}})\right)}, \ \ \ $if$ \ \ \  i=1,2.
\newline\\
\newline\\
\displaystyle  & \frac{\left(-1\right)^{i+1}}{\varepsilon_{3\sigma}({\bf{k}})-\varepsilon_{4\sigma}({\bf{k}})}\prod_{j=1,2}\frac{{\cal{P}}^{\left(2\right)}\left(\varepsilon_{i\sigma}({\bf{k}})\right)}{\left(\varepsilon_{i\sigma}({\bf{k}})-\varepsilon_{j\sigma}({\bf{k}})\right)}, \ \ \ $if$ \ \ \  i=3,4
\end{array}\right.
\label{Equation_A_4}
\end{eqnarray}
with ${\cal{P}}^{\left(2\right)}\left(x\right)$ defined as
\begin{eqnarray}
	&&{\cal{P}}^{(2)}\left(x\right)=a_{2\sigma}x^{2}+b_{2\sigma}x+c_{2\sigma},
	\label{Equation_A_5}
\end{eqnarray}
where
\begin{eqnarray}
	a_{2\sigma}&=&-2\sigma\Delta_{\rm AFM}-\bar{\mu}_{1\sigma}+\bar{\mu}_{2\sigma},
	\nonumber\\
	b_{2\sigma}&=&2\sigma\left(\bar{\mu}_{1\sigma}+\bar{\mu}_{2\sigma}\right)\Delta_{\rm AFM}+\bar{\mu}^{2}_{1\sigma}-\bar{\mu}^{2}_{2\sigma},
	\nonumber\\
	c_{2\sigma}&=&2\Delta^{3}_{\rm AFM}-\left(\bar{\mu}_{1\sigma}-\bar{\mu}_{2\sigma}\right)|\tilde{\gamma}_{\bf{k}}|^{2}-\left(\bar{\mu}_{1\sigma}-\bar{\mu}_{2\sigma}\right)\Delta^{2}_{\rm AFM}
	\nonumber\\
	&&-\bar{\mu}_{1\sigma}\bar{\mu}_{2\sigma}\left(\bar{\mu}_{1\sigma}-\bar{\mu}_{2\sigma}\right)-\left(\bar{\mu}^{2}_{1\sigma}+\bar{\mu}^{2}_{2\sigma}\right)\Delta_{\rm AFM}
	\nonumber\\
	&&+2|\tilde{\gamma}_{\bf{k}}|^{2}\Delta_{\rm AFM}.
	\label{Equation_A_6}
\end{eqnarray}
Furthermore, we get for the coefficient $\gamma_{i{\bf{k}}\sigma}$ in the third equation, in the Eq.(\ref{Equation_22})
\begin{eqnarray}
\footnotesize
\arraycolsep=0pt
\medmuskip = 0mu
\gamma_{i{\bf{k}}\sigma}
=\left\{
\begin{array}{cc}
\displaystyle  & \frac{\left(-1\right)^{i+1}}{\varepsilon_{1\sigma}({\bf{k}})-\varepsilon_{2\sigma}({\bf{k}})}\prod_{j=3,4}\frac{{{\cal{P}}'}^{\left(2\right)}\left(\varepsilon_{i\sigma}({\bf{k}})\right)}{\left(\varepsilon_{i\sigma}({\bf{k}})-\varepsilon_{j\sigma}({\bf{k}})\right)}, \ \ \ $if$ \ \ \  i=1,2.
\newline\\
\newline\\
\displaystyle  & \frac{\left(-1\right)^{i+1}}{\varepsilon_{3\sigma}({\bf{k}})-\varepsilon_{4\sigma}({\bf{k}})}\prod_{j=1,2}\frac{{{\cal{P}}'}^{\left(2\right)}\left(\varepsilon_{i\sigma}({\bf{k}})\right)}{\left(\varepsilon_{i\sigma}({\bf{k}})-\varepsilon_{j\sigma}({\bf{k}})\right)}, \ \ \ $if$ \ \ \  i=3,4,
\end{array}\right.
\label{Equation_A_7}
\end{eqnarray}
where ${{\cal{P}}'}^{\left(2\right)}\left(x\right)$ is 
\begin{eqnarray}
	&&{{\cal{P}}'}^{(2)}\left(x\right)=a_{3\sigma}x^{2}+b_{3\sigma}x+c_{3\sigma}
	\label{Equation_A_8}
\end{eqnarray}
with
\begin{eqnarray}
	&&a_{3\sigma}=2\Delta_{\rm AFM},
	\nonumber\\
	&&b_{3\sigma}=4\mu_{2}\Delta_{\rm AFM},
	\nonumber\\
	&&c_{3\sigma}=2\mu^{2}_{2}\Delta_{\rm AFM}-2\Delta^{3}_{\rm AFM}-2|\tilde{\gamma}_{{\bf{k}}}|^{2}\Delta_{\rm AFM}.
	\label{Equation_A_9}
\end{eqnarray}
The coefficient $\delta_{{\bf{k}}\sigma}$, in the last equation in the Eq.(\ref{Equation_22}), was obtained as
\begin{eqnarray}
\footnotesize
\arraycolsep=0pt
\medmuskip = 0mu
\delta_{i{\bf{k}}\sigma}
=\left\{
\begin{array}{cc}
\displaystyle  & \frac{\left(-1\right)^{i+1}}{\varepsilon_{1\sigma}({\bf{k}})-\varepsilon_{2\sigma}({\bf{k}})}\prod_{j=3,4}\frac{{\cal{P}}^{(2)}_{\Delta}\left(\varepsilon_{i\sigma}({\bf{k}})\right)}{\left(\varepsilon_{i\sigma}({\bf{k}})-\varepsilon_{j\sigma}({\bf{k}})\right)}, \ \ \ $if$ \ \ \  i=1,2.
\newline\\
\newline\\
	\displaystyle  & \frac{\left(-1\right)^{i+1}}{\varepsilon_{3\sigma}({\bf{k}})-\varepsilon_{4\sigma}({\bf{k}})}\prod_{j=1,2}\frac{{\cal{P}}^{(2)}_{\Delta}\left(\varepsilon_{i\sigma}({\bf{k}})\right)}{\left(\varepsilon_{i\sigma}({\bf{k}})-\varepsilon_{j\sigma}({\bf{k}})\right)}, \ \ \ $if$ \ \ \  i=3,4,
\end{array}\right.
\label{Equation_A_10}
\end{eqnarray}
where
\begin{eqnarray}
	&&{\cal{P}}^{(2)}_{\Delta}\left(x\right)=-\left(\Delta_{\sigma}+\gamma_1\right)\left[x^{2}-x\left(\bar{\mu}_{1\sigma}+\bar{\mu}_{2\sigma}\right)\right.
	\nonumber\\
	&&\left.-\Delta^{2}_{\rm AFM}+\sigma\left(\bar{\mu}_{1\sigma}-\bar{\mu}_{2\sigma}\right)\Delta_{\rm AFM}+\bar{\mu}_{1\sigma}\bar{\mu}_{2\sigma}\right].
	\nonumber\\
	\label{Equation_A_11}
\end{eqnarray}
Let's mention at the end that the coefficients $\alpha_{i{\bf{k}}\sigma}$, $\beta_{i{\bf{k}}\sigma}$
$\gamma_{i{\bf{k}}\sigma}$ and $\delta_{i{\bf{k}}\sigma}$ in the Eqs.(\ref{Equation_A_1}), (\ref{Equation_A_4}), (\ref{Equation_A_7}) and (\ref{Equation_A_10}), are dimensionless due to the structure of the polynomials given above. 
\section*{References}
\bibliography{references_authors}

\end{document}